\documentclass[graybox,envcountchap,sectrefs]{svmono}

\usepackage{mathptmx}
\usepackage{helvet}
\usepackage{courier}
\usepackage{type1cm}
\usepackage[round, sort, numbers, authoryear]{natbib}
\usepackage{makeidx}         
\usepackage{graphicx}        
\usepackage{multicol}        
\usepackage[bottom]{footmisc}

\makeindex             

\begin{document}

\author{Roland Diehl, Dieter H. Hartmann, and Nikos Prantzos (Editors)}
\title{Lecture Notes in Physics 812 - Astronomy with Radioactivities}
\subtitle{(An Introduction to Astrophysics with Decaying Isotopes)}
\maketitle

\frontmatter

\tableofcontents

\mainmatter

%
%
\let\jnl=\rmfamily
\def\refe@jnl#1{{\jnl#1}}%

\newcommand\aj{\refe@jnl{AJ}}%
\newcommand\actaa{\refe@jnl{Acta Astron.}}%
\newcommand\araa{\refe@jnl{ARA\&A}}%
\newcommand\apj{\refe@jnl{ApJ}}%
\newcommand\apjl{\refe@jnl{ApJ}}%
\newcommand\apjs{\refe@jnl{ApJS}}%
\newcommand\ao{\refe@jnl{Appl.~Opt.}}%
\newcommand\apss{\refe@jnl{Ap\&SS}}%
\newcommand\aap{\refe@jnl{A\&A}}%
\newcommand\aapr{\refe@jnl{A\&A~Rev.}}%
\newcommand\aaps{\refe@jnl{A\&AS}}%
\newcommand\azh{\refe@jnl{AZh}}%
\newcommand\gca{\refe@jnl{GeoCh.Act}}%
\newcommand\memras{\refe@jnl{MmRAS}}%
\newcommand\mnras{\refe@jnl{MNRAS}}%
\newcommand\na{\refe@jnl{New A}}%
\newcommand\nar{\refe@jnl{New A Rev.}}%
\newcommand\pra{\refe@jnl{Phys.~Rev.~A}}%
\newcommand\prb{\refe@jnl{Phys.~Rev.~B}}%
\newcommand\prc{\refe@jnl{Phys.~Rev.~C}}%
\newcommand\prd{\refe@jnl{Phys.~Rev.~D}}%
\newcommand\pre{\refe@jnl{Phys.~Rev.~E}}%
\newcommand\prl{\refe@jnl{Phys.~Rev.~Lett.}}%
\newcommand\pasa{\refe@jnl{PASA}}%
\newcommand\pasp{\refe@jnl{PASP}}%
\newcommand\pasj{\refe@jnl{PASJ}}%
\newcommand\skytel{\refe@jnl{S\&T}}%
\newcommand\solphys{\refe@jnl{Sol.~Phys.}}%
\newcommand\sovast{\refe@jnl{Soviet~Ast.}}%
\newcommand\ssr{\refe@jnl{Space~Sci.~Rev.}}%
\newcommand\nat{\refe@jnl{Nature}}%
\newcommand\iaucirc{\refe@jnl{IAU~Circ.}}%
\newcommand\aplett{\refe@jnl{Astrophys.~Lett.}}%
\newcommand\apspr{\refe@jnl{Astrophys.~Space~Phys.~Res.}}%
\newcommand\nphysa{\refe@jnl{Nucl.~Phys.~A}}%
\newcommand\physrep{\refe@jnl{Phys.~Rep.}}%
\newcommand\procspie{\refe@jnl{Proc.~SPIE}}%

\newcommand{\xspace}{~}
\newcommand{\Al}{$^{26}$Al\xspace}
\newcommand{\al}{$^{26}$Al}
\newcommand{\Be}{$^{7}$Be\xspace}
\newcommand{\be}{$^{7}$Be}
\newcommand{\bem}{$^{10}$Be\xspace}
\newcommand{\ca}{$^{44}$Ca}
\newcommand{\Ca}{$^{44}$Ca\xspace}
\newcommand{\cam}{$^{41}$Ca\xspace}
\newcommand{\Co}{$^{56}$Co\xspace}
\newcommand{\co}{$^{56}$Co\xspace}
\newcommand{\csm}{$^{135}$Cs\xspace}
\newcommand{\cs}{$^{135}$Cs}
\newcommand{\ct}{$^{13}$C}
\newcommand{\ci}{$^{57}$Co\xspace}
\newcommand{\Ci}{$^{57}$Co\xspace}
\newcommand{\ch}{$^{60}$Co\xspace}
\newcommand{\Ch}{$^{60}$Co\xspace}
\newcommand{\Cl}{$^{36}$Cl\xspace}
\newcommand{\cl}{$^{36}$Cl}
\newcommand{\li}{$^{7}$Li}
\newcommand{\Li}{$^{7}$Li\xspace}
\newcommand{\Fe}{$^{60}$Fe\xspace}
\newcommand{\fh}{$^{60}$Fe\xspace}
\newcommand{\fe}{$^{56}$Fe}
\newcommand{\Fr}{$^{57}$Fe\xspace}
\newcommand{\fr}{$^{57}$Fe\xspace}
\newcommand{\mg}{$^{26}$Mg}
\newcommand{\Mg}{$^{26}$Mg\xspace}
\newcommand{\mn}{$^{54}$Mn}
\newcommand{\Na}{$^{22}$Na\xspace}
\newcommand{\Ne}{$^{22}$Ne\xspace}
\newcommand{\Ni}{$^{56}$Ni\xspace}
\newcommand{\nh}{$^{60}$Ni\xspace}
\newcommand{\Nh}{$^{60}$Ni\xspace}
\newcommand\nuk[2]{$\rm ^{\rm #2} #1$}  
\newcommand{\pd}{$^{107}$Pd\xspace}
\newcommand{\pb}{$^{205}$Pb}
\newcommand{\tc}{$^{99}$Tc}
\newcommand{\Sc}{$^{44}$Sc\xspace}
\newcommand{\Ti}{$^{44}$Ti\xspace}
\newcommand{\ti}{$^{44}$Ti\xspace}
\def\aa{$\alpha$}
\newcommand{\about}{$\simeq$}
\newcommand{\cms}{cm\ensuremath{^{-2}} s\ensuremath{^{-1}}\xspace}
\newcommand{\degree}{$^{\circ}$}
\newcommand{\flux}{ph~cm\ensuremath{^{-2}} s\ensuremath{^{-1}}\xspace}
\newcommand{\fluxrad}{ph~cm$^{-2}$s$^{-1}$rad$^{-1}$\ }
\newcommand{\ga}{\ensuremath{\gamma}}
\newcommand{\gam}{\ensuremath{\gamma}}
\def\nn{$\nu$}
\def\ra{$\rightarrow$}
\newcommand{\Msol}{M\ensuremath{_\odot}\xspace}
\newcommand{\msol}{M\ensuremath{_\odot}\xspace}
\newcommand{\ms}{M\ensuremath{_\odot}\xspace}
\newcommand{\msb}{M\ensuremath{_\odot}\xspace}
\newcommand{\Msun}{M\ensuremath{_\odot}\xspace}
\def\mspcb{M$_{\odot}pc^{-2}$}
\def\msy{M$_{\odot}y^{-1}$}
\newcommand{\Rsun}{R\ensuremath{_\odot}\xspace}
\newcommand{\rsun}{R\ensuremath{_\odot}\xspace}
\newcommand{\Lsun}{L\ensuremath{_\odot}\xspace}
\newcommand{\lsun}{L\ensuremath{_\odot}\xspace}
\newcommand{\solar}{\ensuremath{_\odot}\xspace}
\newcommand{\zs}{Z\ensuremath{_\odot}\xspace}





%
%
%
%
 \chapauthor{Maurizio Busso}
\chapter{The early Solar System}

\label{chap:6}

This chapter presents a (partial) review of the information
we can derive on the early history of the Solar System from radioactive nuclei of
very different half-life, which were recognized to have been present alive in pristine solids. In fact, radioactivities open for us a unique window on the evolution of the
solar nebula and provide tools for understanding the crucial events that determined
and accompanied the formation of the Sun. Discussing these topics will require consideration of (at least) the following issues. i) The determination of an age for solar system bodies, as it emerged especially from the application of radioactive dating. ii) A synthetic account of the measurements that proved the presence of radioactive nuclei (especially those of half-life lower than about 100 Myr) in the Early Solar System (hereafter ESS). iii) An explanation of their existence in
terms of galactic nucleosynthesis, and/or of local processes (either exotic or
in-situ) preceding and accompanying the formation of the Sun. This will also need some
reference to the present scenarios for star formation, as applied to the ESS.


\section{The Age of the Solar System}
\label{sec:6.1}
In the second half of the XIXth century, Helmholtz and Kelvin
independently noticed that, evolving under the opposite effects of
gravity and thermodynamics, the Sun would survive only for about 20
Million years (the \emph{Kelvin-Helmholtz time scale}).
Hints on the fact that this estimate was too short first came from
Charles Darwin and his disciples, as this datum appeared to be by far
insufficient to permit the biological and geological evolution they
were discovering.

The first role played by radioactivities in the history of our
understanding of the solar system was then the solution of this
dilemma concerning its age. This solution became available rather soon, less
than two decades after the same discovery of radioactivity as a physical phenomenon.

\subsection{The Beginnings}
\label{sec:6.1.1}
In 1895 Wilhelm Conrad R\"ontgen showed the amazing imaging properties of the
energetic radiation emanating (sometimes) from matter; the X-rays.
Soon after Becquerel made the fortuitous discovery that uranium
produced X-rays leaving traces on photographic plates. Further
researches on these "beams" were carried out by Pierre and Marie
Curie. In 1898 they showed that pitchblende (the major uranium ore,
mainly made of uranium oxide UO$_2$), emitted X-rays 300 times
stronger than those expected from U-rich compounds. This required
the presence of another X-ray emitter, a nucleus that was called
polonium (from Marie's native land). Then they discovered radium,
inventing the term "radio-activity", and showed that the X-ray
emission could be quantitatively predicted as a function of time.

It was then Rutherford who unambiguously showed how radioactivity
were an intrinsic property of certain atoms, linked to their
intimate structure; and Arthur Holmes, in 1911, presented the first
systematic attempt at dating rocks, based on their content of U and
Pb. The ages of the rocks derived from radioactivity were scattered
from 0.64 to more than 1.4 Gyr, in sharp contrast with all
non-nuclear estimates. The decades to follow would prove that
Holmes was right: the age of the solar
system has now been suggested to be 4.566 $\pm$ 0.001 Gyr, see
e.g. \citet{all}.

\citet{ruth} also
demonstrated that the ratio $^{235}$U/$^{238}$U (0.007 presently)
was probably as high as 0.3 at the Sun's formation, and concluded
that an extrapolation backward would yield the production ratio in
the extra-solar environment where nuclei had been produced. So it was
Rutherford who first suggested to use radioactivity for dating not
only planetary rocks and the Sun, but even the environment from where the Solar System
was formed; and he correctly estimated that U production was already going on
4$-$5 Gyrs ago, although at his epoch the age of the Sun was still
largely debated. Most of the tools for dating samples
with radioactive nuclei had been, at this point, made available.

Rutherford's and Holmes' suggestion of using radioactivities as
clocks capable of dating ancient samples remains as one of the most
important scientific results of the XXth century. Depending on the
isotope used and its lifetime, the technique is good for measuring
very different ages: from those of historical and archaeological
objects to those of materials in the ESS and in old stars of the
Galaxy. These tools are generally known under the name of
``radio-chronology'' or, in geological and astronomical contexts,
``nuclear cosmo-chronology'' ~\citep{clay88}. For a detailed history of
the Earth's datation see \citet{darl}; for an account of more modern
efforts, after Holmes' work, see Chapter 2 in this
volume.
\subsection{Long-Lived Nuclei for Solar System Dating}
\label{sec:6.1.2}
Estimates of the solar-system age are now based on the
measured abundances of long-lived radioactive (parent) nuclei and
of their stable daughters in pristine meteorites. This kind of
radiometric dating can be performed through modern
upgrades of the mass spectrometer, from tiny samples of ancient
solid materials, making use of unstable
isotopes preserving today a residual abundance.  This implies that
their lifetime must be very long; and, actually, the
shortest-lived nucleus fulfilling this requirement is $^{235}$U
($\tau = 1.015\cdot10^9$ yr).

\begin{figure}[ht!!]
\begin{center}
\includegraphics[width=0.8\columnwidth,clip]{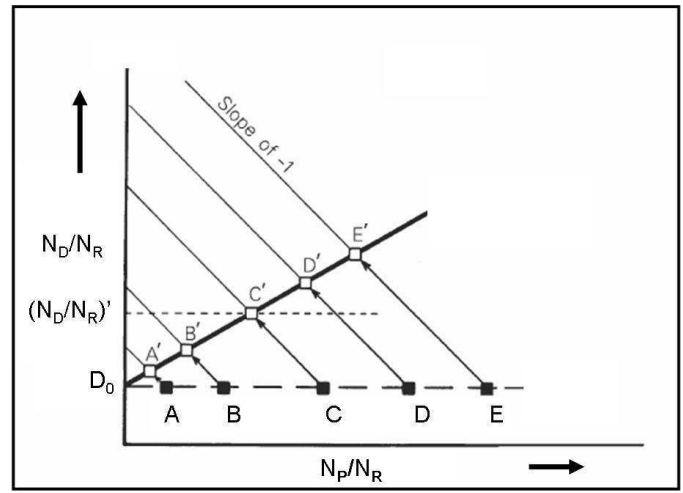}
\vskip 0pt \caption{The scheme for the construction of an
isochron in the decay of an unstable parent to its daughter.
See text for comments.} \label{fig:one}
\end{center}
\end{figure}

\begin{table}[ht]
\begin{center}
{\bf Table 1. Possible decay modes at atomic mass $A$ and
charge $Z$}
\vspace{0.5cm}
\begin{tabular}{|l|l|l|}
\hline
Decay Mode & Particles involved & Daughter   \\
           &                    &  nucleus   \\
\hline
$\beta^{-}$-decay      & emission of an e$^{-}$ and $\bar \nu$       & (A, Z+1)   \\
\hline
$\beta^{+}$-decay      & emission of an e$^{+}$ and a $\nu$          & (A, Z-1)   \\
\hline
e$^{-}$-capture        & e$^{-}$ capture, $\nu$ emission & (A, Z-1)   \\
\hline
double $\beta^{-}$-decay & emission of two e$^{-}$ and two $\bar \nu$s      & (A, Z+2)   \\
\hline
double e$^{-}$-capture & capture of two e$^{-}$, two $\nu$ emissions & (A, Z-2) \\
\hline
e$^{-}$-capture, e$^{+}$-emission & capture of an e$^{-}$, emission & (A, Z-2) \\
                              & of an e$^{+}$ and two $\nu$s             &          \\
\hline
double e$^{+}$ emission & emission of two e$^{+}$ and two $\nu$s           & (A, Z-2) \\
\hline
 $\alpha$-decay      & an $\alpha$ particle (A=4, Z=2) is emitted    & (A-4, Z-2)  \\
\hline
 proton emission     & a proton is ejected                           & (A-1, Z-1)  \\
\hline
 neutron emission    & a neutron is emitted                          & (A-1, Z)    \\
\hline
 spontaneous fission & two or more smaller nuclei emitted    &             \\
\hline
 cluster decay       & emission of a nucleus of A1, Z1 & (A-A1,Z-Z1)\\
                     &                                       & + (A1, Z1) \\
\hline
$\gamma$-decay       & photon emission from excited states  & (A, Z) \\
\hline
\end{tabular}
\end{center}
\label{tab1}
\end{table}

Various parent-daughter couples can be used, exploiting the existing channels for
natural decay (schematically summarized in Table 1). General
references where these techniques are discussed in an astrophysical
environment are numerous: see e.g. Chapter 2 in this volume,
and \citet{clay88,pag,cow,darl}.

With reference to Figure \ref{fig:one}, the age calculation
takes into account the presence of the radioactive parent ($P$)
and also the original amount of the stable daughter ($D$) at
the beginning of the time interval. In general,
when there is a non-radiogenic isotope of the daughter element
in the mineral, it can be used as a reference to compute
abundance ratios (in this case this reference nucleus is indicated
as $R$). Now suppose a series of rocks are characterized, at formation, by the values
$A$, $B$, ... $E$ of the ratio $N_P/N_R$; they all obviously have
the same {\it initial} ($N_D/N_R$)$_0$ = $D_0$.

After a
time interval $t$, the content of $P$ will be diminished in each sample,
so that, for example:
$$
(N_P/N_R)^A_t = (N_P/N_R)^A_0 \exp({-\lambda t})
$$
where $\lambda$ is the decay constant ($\lambda$ =
0.6931/$t_{1/2}$). As the product $\lambda t$ is the same for all rocks,
the new contents of $P$ (abscissas of the measured points $A^{'}, B^{'},...$) are
proportional to the original ones and the measurements themselves lay on
a straight line (isochron). At this same moment the value of $N_D/N_R$ is the
sum of the radiogenic contribution and of the initial value $D_0$. It is
straightforward that:
$$
(N_D/N_R)_t = D_0 + (N_P/N_R)_t [\exp({\lambda t})
- 1] \eqno(1)
$$
\begin{figure*}[ht!!] 
\begin{center}
\includegraphics[height=12cm,angle=-90]{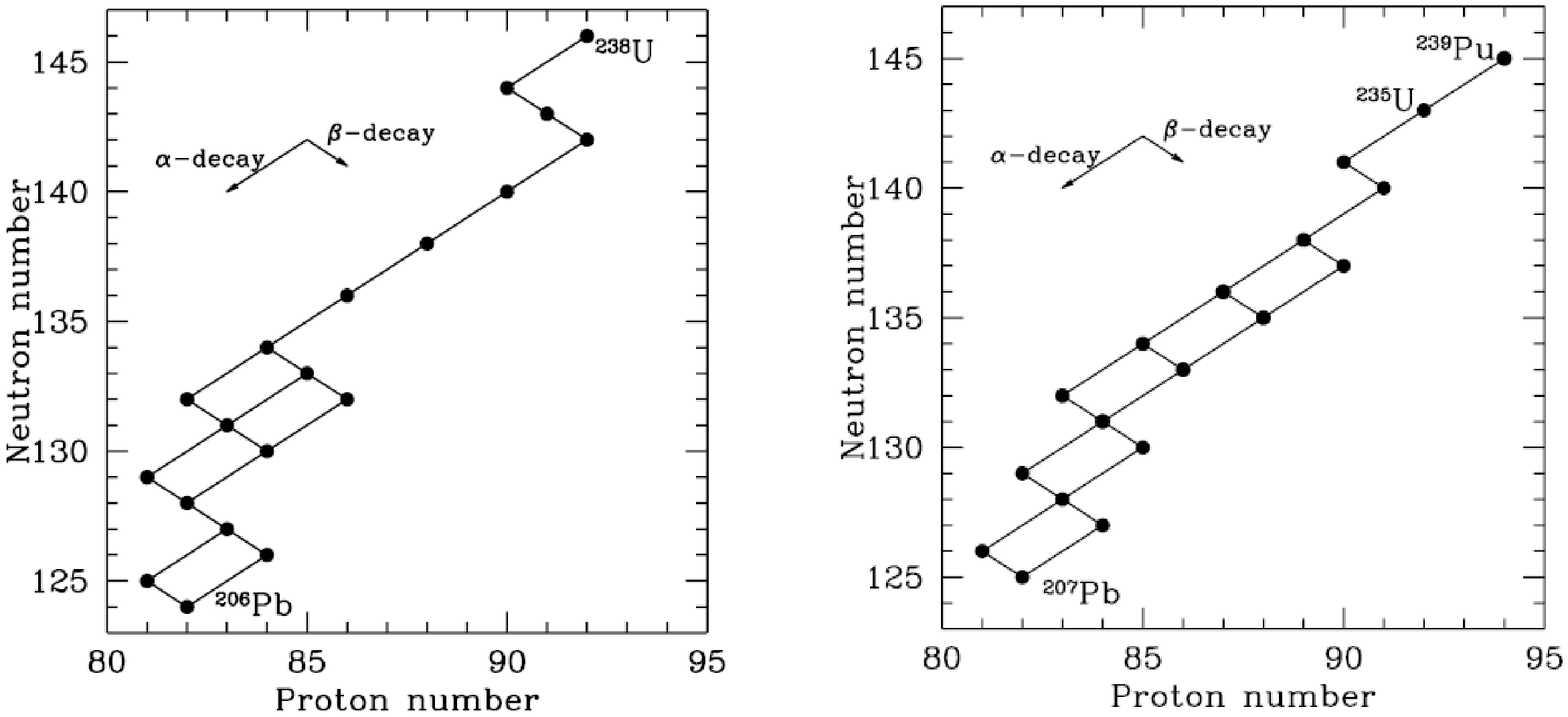}
\end{center}
\vskip 0pt \caption{The decay chains of $^{238}$U and $^{235}$U, commonly used
for dating rocks. The U-Pb method is applied to the mineral zircon
(ZrSiO$_4$), containing some U (because it can chemically
substitute Zr) and strongly rejecting lead}\label{fig:two}
\end{figure*} 

When measuring rocks with the same age $\Delta t$, the isochron passing through
the points $A^{'}, B^{'},...$ will have an intercept
on the ordinate axis providing the initial isotopic ratio of the daughter, $D_0$.
A least square technique then allows one to determine both the age and
the initial $D_0$ ratio. For a radiogenic nucleus $R$ decaying to a stable isotope $D$, with a half-life $t_{1/2}$, it is immediate to derive that
$$
\Delta t = {1\over \lambda} \times ln \left[1+{{\Delta D }\over{\Delta R}}\right]
\eqno(2)$$
A recent example of age determination with the
described technique, involving $^{87}$Rb ($P$), $^{87}$Sr ($D$) and
$^{86}$Sr ($R$) can be found in \citet{ander}

Several parent-daughter couples of interest for isotope geology exist: for example the $^{40}$K $-$ $^{40}$Ar pair ($^{40}$K producing $^{40}$Ar through $\beta^{+}$-decay and $e^{-}$-captures, with a half-life of 1.25 Gyr) and the $^{147}$Sm $-$ $^{143}$Nd pair
(the first $\alpha$-decays to the second with a half life of 10.6 Gyr).

Alternatively, the series of $\alpha$ and $\beta$ decays of
uranium (and sometimes transuranic nuclei) can be exploited.
Perhaps the most commonly-used pair is in this case U-Pb, actually
made of two separate chains, one leading from $^{238}$U to
$^{206}$Pb, with a half-life of 4.5 Gyr, the second leading from
$^{235}$U to $^{207}$Pb, with a half-life of 0.7 Gyr. The two
processes are shown in Figure \ref{fig:two}.
Very recently, cautions have been advanced on the accuracy of the U-Pb datation,
due to the discovery of live $^{247}$Cm in the ESS. This nucleus decays to $^{235}$U, with a half-life of 15.6 Myr, so that the initial inventory of U must be corrected for this effect in order to avoid inaccuracies in the solar system age, which can be as high as 5 Myr \citep{bren1}. For a more detailed account of Solar-System datation with long-lived nuclei, see again Chapter 2 in this volume.

\section{Short-lived Radioactive Nuclei in the ESS}
\label{sec:6.2}
\begin{figure}[t!!]
\begin{center}
\includegraphics[width=0.8\columnwidth,clip]{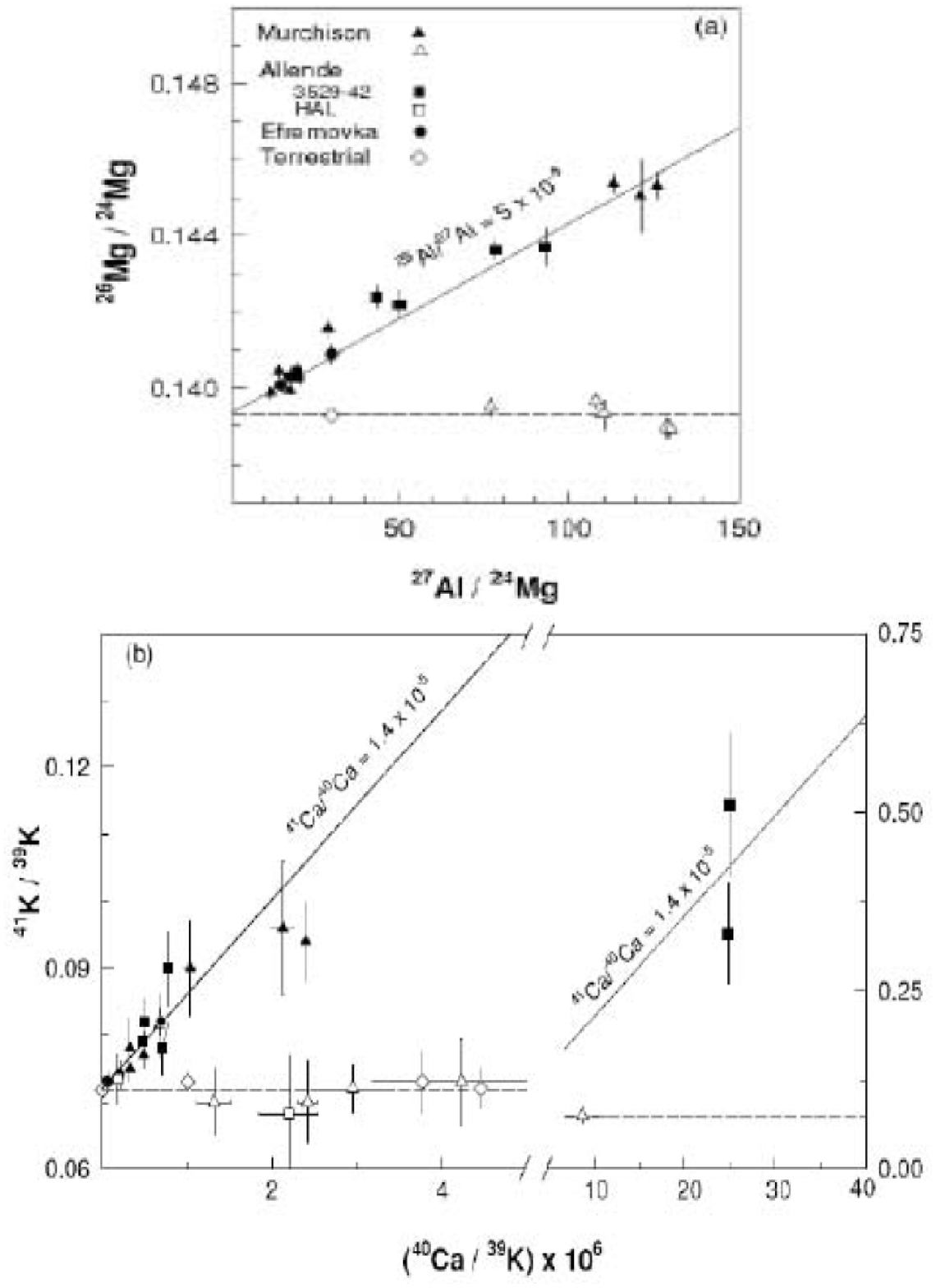}
\vskip 0pt \caption{Panel a) shows the Al-Mg data from CAIs in different meteorites; filled symbols demonstrate the presence in ESS of $^{26}$Al at the level $^{26}$Al/$^{27}$Al = $5\cdot10^{-5}$. Panel b) displays in a similar way the Ca-K data from the same samples. The first published data, and the line showing [$^{41}$Ca/$^{40}$Ca]$_0$ = 1.4$\cdot$10$^{-8}$ were due to \citet{sug94} and \citet{sri96}. Other data and the demonstration of the correlation with Al are from \citet{sahij}. The figure is adapted from these last authors and from \citet{was06}. Copyright Nature Publishing Group.}
\label{fig:three}
\end{center}
\end{figure}

The measurements revealing the presence, in the ESS, of radioactive
species with meanlife lower than about 100 Myr (generally referred to as "short-lived") were the source of an enormous progress in our knowledge of the solar origins and have by now been recognized as one of the most important scientific achievements in the second half of the XXth century. The identification of the decay product allows us to
draw a plot similar to Figure \ref{fig:one}, if we replace in the abscissa the parent nucleus P (that no longer exists) with another stable isotope of the same element (sometimes called the "substitute", or "index" nucleus) supposedly entered into the sample together with P \citep{bus99,was06}. In this way, after a careful demonstration that the excess in the stable daughter is quantitatively correlated with the chemical properties of the parent element (using the index isotope), one can reconstruct isochrones and derive the initial concentration of P in the sample. In a diagram similar to Figure \ref{fig:one}, and using the couple $^{26}$Al $-$ $^{26}$Mg as an example, this requires
to use on the abscissa the ratio $^{27}$Al/$^{24}$Mg ($^{27}$Al is then
the index nucleus) and on the ordinate
$^{26}$Mg/$^{24}$Mg. This can be done, for example, for a mineral
intrinsically poor in Mg, where the presence of $^{26}$Mg can be
ascribed largely to the $^{26}$Al-decay. A description of the procedure can be found in ~\citet{lee}. Figure \ref{fig:three} shows the technique applied to $^{26}$Al and $^{41}$Ca, as obtained by \citet{sahij}. These authors showed that $^{26}$Al and $^{41}$Ca were either both present and correlated, or both absent from the samples, thus demonstrating their probable common origin.

\begin{table}[h!!]
\begin{center}
{\bf Table 2. Short-Lived Nuclei in the ESS (*)}

\vspace{0.5cm}

\begin{tabular}{|c|c|c|c|}
\hline
  Rad. & Ref. & Mean Life & $(N_P/N_I)_{ESS}$   \\
       &      &    (Myr)  &                      \\
\hline
  $^{10}$Be & $^{9}$Be  & 2.2 & $5.2\cdot10^{-4}$  \\
  $^{26}$Al & $^{27}$Al  & 1.05 & $5\cdot10^{-5}$  \\
  $^{36}$Cl & $^{35}$Cl  & 0.43 &  $10^{-4}$   \\
  $^{41}$Ca & $^{40}$Ca  & 0.15  & $\ge 1.5\cdot10^{-8}$   \\
  $^{53}$Mn & $^{55}$Mn  & 5.3 & $6.7\cdot10^{-5}$   \\
  $^{60}$Fe & $^{56}$Fe  & 2.2(**) & $10^{-7} - 10^{-6}$ \\
  $^{107}$Pd & $^{108}$Pd & 9.4 & $2.0\cdot10^{-5}$   \\
  $^{129}$I & $^{127}$I  & 23 & $1.0\cdot10^{-4}$   \\
  $^{146}$Sm & $^{144}$Sm & 148 & $1.0\cdot10^{-2}$ \\
  $^{182}$Hf & $^{180}$Hf  & 13 & $2.0\cdot10^{-4}$ \\
  $^{244}$Pu & $^{232}$Th  & 115 & $3\cdot10^{-3}$ \\
  $^{247}$Cm & $^{232}$Th  & 23 & $1$-$2.4\cdot10^{-4}$\\
\hline
 $^{135}$Cs & $^{133}$Cs  & 2.9 & $1.6\cdot10^{-4}$ (?)\\
 $^{205}$Pb &  $^{204}$Pb & 22 & $1 - 2\cdot10^{-4}$ (?)\\

\hline
\end{tabular}
\end{center}
(*) In the lower panel we include isotopes for which only hints
or upper bounds are available. [For references see \citet{was06} and, for $^{247}$Cm, \citet{bren2}]\\
(**) This is the value used in most works. Recent measurements by \citet{rugel}
suggest intead 3.8 Myr.
\label{tab2}
\end{table}

The mentioned measurements revealed a previously unsuspected complexity for
the physical processes that affected the early phases of the solar system, linking
stellar nucleosynthesis, galactic chemical evolution and the energetic processes
in the winds of the early Sun (see also Chapters 2 and 7 in this volume). These discoveries started with the work by \citet{rey}, who was the first to show that a now extinct nucleus, $^{129}$I ($\tau$ = 23 Myr), had existed alive in the ESS. This was demonstrated analyzing the concentration of its decay product,
$^{129}$Xe, in meteorites \citep{jr61}. It was then suggested by  ~\citet{was60} and by \citet{cam60} that the abundance of
$^{129}$I could be attributed to the continuous production of
$r$-process nuclei in the galaxy, provided the solar nebula had
been isolated for about 10$^8$ years.

Shortly after the discovery of iodine, the presence of another relatively long-lived
nucleus, $^{244}$Pu ($\tau$ = 115 Myr), was inferred from excesses, in planetary differentiates (PD), of neutron-rich Xe isotopes \citep{row65}.
It was then demonstrated that the enrichment in Xe isotopes correlated with
excess fission tracks in meteoritic materials rich in both U, Th, and rare earth
elements, showing that fission of heavy nuclei had occurred  in situ \citep{was69,sp77}

More systematic measurements, extending to nuclei of shorter
lifetime and to very early ESS constituents, became possible
after 1969, thanks to the fall of two extremely primitive meteorites: one
in the Mexican village Allende and the second in the Australian village Murchison.
They are rich in very refractory aggregates, called ``Aluminium-Calcium Inclusions'', or CAIs \citep{map03}, which are believed to be among the first condensed solids in the solar nebula.

Measurements of Xe isotopes on CAIs in Allende \citep{pl72} soon confirmed both the presence
of $^{129}$I and the in situ fission of actinides. Even more important was the discovery of
$^{26}$Mg excesses that, thanks to the technique illustrated above in Figure \ref{fig:one} and in Figure \ref{fig:three}, were unambiguously proven to be the relics of the earlier presence
of $^{26}$Al \citep{lee,lee1}. The meanlife of $^{26}$Al ($\tau = 1.05 $ Myr) is much shorter
than for iodine, and its presence seemed to confirm older suggestions \citep{urey,ud56},
where it had been anticipated that $^{26}$Al would be a very likely source for the early heating
of planetesimals, as other internal energy sources appeared insufficient to melt them. Searches
for $^{26}$Al had been done before the fall of Allende \citep{sch1}, but they could not find the
nucleus in solid solar-system samples. After the works by \citet{lee,lee1}, several studies then
confirmed the homogeneous presence of $^{26}$Al at the high (often called {\it canonical}) ratio
$^{26}$Al/$^{27}$Al = 5$\times$10$^{-5}$. The existence of values higher than the standard one (sometimes referred to as "super canonical") was suggested a few years ago \citep{y+05,liu}, but not confirmed afterwards.

The above discoveries were soon followed by others: the
$p$-nucleus $^{146}$Sm ($\tau = 148$ Myr) was identified in meteoritic materials more
evolved than CAIs, showing the signs of ongoing planetary
differentiation: it was identified from its decay product, $^{142}$Nd \citep{lug77}.
Then the abundance of $^{107}$Pd ($\tau = 9.4$ Myr) was
established, again in PDs \citep{kel}. The fact that planetary cores
could contain a relatively short-lived nucleus like $^{107}$Pd was a proof
that they formed very quickly in the history of the solar system.

Excesses of $^{205}$Tl were also inferred by \citet{chw}. They would
indicate the presence of $^{205}$Pb ($\tau$ = 22 Myr). Recently \citet{niel}
found evidence of a correlation between the $^{205}$Tl excess in iron
meteorites and  $^{204}$Pb. This would imply the in-situ decay
of $^{205}$Pb, at the level $^{205}$Pb/$^{204}$Pb = 1-2$\times$10$^{-4}$.
We notice that $^{205}$Pb is a shielded (``s-process only'') n-capture nucleus,
whose presence would certify that the ESS was somehow influenced by recent $s$-process nucleosynthesis events, either from high mass stars \citep{bg85} or from low-intermediate mass red giants \citep{gal8}; see also later, in section 6.4.3.

The presence of $^{53}$Mn ($\tau$ =5.3 Myr) in CAIs, through a correlation of $^{53}$Cr
with Mn, was suggested by \citet{bir85,bir88}. This was later confirmed by
measurements of abundant $^{53}$Mn in PDs, made first by \citet{lms92} and subsequently by \citet{ls98} and \citet{hkk98}.  $^{53}$Mn is very effectively
produced in supernovae (SNe II or SNIa), and is also a possible outcome of spallation
processes.

Tungsten isotopic anomalies discovered in early planetary materials, with deficiencies in $^{182}$W, and their correlation with hafnium in chondritic meteorites demonstrated the presence of $^{182}$Hf ($\tau$ = 13 Myr), see \citet{lee2,lee3,hj96}.

The presence of  $^{60}$Fe ($\tau$ =2.2 Myr) was early discovered by \citet{sl31,sl32}. Subsequently $^{60}$Ni excesses were shown to be correlated with Fe/Ni ratios in chondrites,
and the $^{60}$Fe concentration was found to be rather high in the ESS \citep{mos1,mos2,th03}. This nucleus is a product of neutron captures at a relatively high neutron density, and can be synthesized both in massive stars exploding as supernovae and in intermediate-mass stars in their final evolutionary stages, but not in spallation processes \citep{lee4}.
Recent estimates of the precise initial $^{60}$Fe content of the ESS have oscillated
noticeably, ranging from a minimum of $^{60}$Fe/$^{56}$Fe = 10$^{-7}$ to a maximum some 20 times higher \citep{wd07}. Despite many discussions and claimed revisions, the situation still remains essentially as defined by \citet{th06}, where a recommended range of (0.5 - 1)$\times$10$^{-6}$ was indicated; see also discussion in \citet{gm10,hus09}. A remarkable suggestion by \citet{b+07}, according to which deficits in $^{60}$Ni would exist in early meteorites as compared to the planets, and would indicate a contamination in $^{60}$Fe that occurred shortly after, not before, the solar nebula formation, seem now to be the result of experimental uncertainties and have not since been confirmed \citep{dau1,dau2}.

$^{10}$Be ($\tau$ =2.3 Myr) was shown to be present in the ESS by \citet{mcr}; as it is not produced by stellar nucleosynthesis, its presence certifies  either the in situ proton bombardment of small solids, or some contributions from galactic cosmic rays. Recent measurements fix the initial $^{10}$Be/$^{9}$Be ratio at
(0.4 $-$ 1)$\times$10$^{-3}$ \citep{liu}.

Also very-short lived nuclei, with lifetimes below 1 Myr, were discovered in the last decade of the XXth century. The presence of $^{41}$Ca ($\tau$ =0.15 Myr) was ascertained by \citet{sug94,sri96}.
The initial ratio derived was $^{41}$Ca/$^{40}$Ca = 1.5$\times$10$^{-8}$, a datum that was later suggested to be a lower limit \citep{gou06}. $^{41}$Ca is abundantly produced from neutron captures in stellar environments, but the short time scale for its decay poses hard constraints on any astrophysical scenario for its formation. At the abundances observed, it might also be produced by proton bombardment, but its proven correlation with $^{26}$Al, already shown in Figure \ref{fig:three} \citep{sahij}, and the recent evidence that $^{26}$Al must be exotic \citep{fit}, indicate that also $^{41}$Ca should come from stellar nucleosynthesis.

Hints on the presence of alive $^{36}$Cl ($\tau$=0.43 Myr) in the ESS were early found by \citet{gb82}, through measurements of $^{36}$Ar abundances in Allende samples (including CAIs). The large shifts found in $^{36}$Ar/$^{38}$Ar correlated with Cl abundances, and would imply an initial ratio $^{36}$Cl/$^{35}$Cl $\simeq$ 2$\times$ 10$^{-8}$ (U. Ott, personal communication). A more recent study
by \citet{mur97} also found $^{36}$Ar shifts that were attributed to $^{36}$Cl. Subsequent works \citep{lin1,lin2} clearly demonstrated a correlation of $^{36}$S/$^{34}$S with $^{35}$Cl/$^{34}$S in late-formed halogen-rich phases in CAIs ($^{36}$S is another decay product of $^{36}$Cl). Sulphur anomalies  were shown to be uncorrelated with $^{26}$Al. These results demonstrate the presence of $^{36}$Cl in the ESS with a high concentration ($^{36}$Cl/$^{35}$Cl $\simeq 10^{-4}$). Stellar sources would not be capable of explaining $^{36}$Cl at such  a high abundances, and it would require to be attributed to bombardment of solids by the early solar wind, while $^{26}$Al would in this case come from stars. These results were subsequently strengthened by the work by \citet{hsu}, where the decoupling of chlorine and aluminium was unambiguously and clearly demonstrated. Further evidence have then been added \citep{jac09,matz10}, confirming that the use of $^{36}$Cl as a chronometer for ESS events is unfeasible and that chlorine was added by phenomena internal to the system, occurred well after the injection of $^{26}$Al.

The wealth of new measurements on ESS samples has now become impressive. The above account is certainly incomplete, but should at least focus our attention on the fact that no simple explanation for the origin of the complex nucleosynthesis pattern revealed by short-lived radioactivities can be invoked. It is evident that, although solid materials formed in a very short lapse of time (from a fraction of a Myr to a few Myrs), they maintain the records of several phenomena, from a blend of different stellar nucleosynthesis processes, to solar wind bombardment and possibly also to galactic cosmic ray spallation.

Table 2 summarizes the status of ESS radioactivities today (for
notations, see later equations 2 and 3). A lively scientific debate is
now considering the various possible sources for the production of
short lived nuclei, including the continuous synthesis in the Galaxy, presented e.g.
in \citet{sch} and in Chapter 7 of this volume; the local pollution by a
nearby star, early introduced by \citet{ct77} for a massive supernova
and recently addressed by \citet{was06} and by \cite{tr09} for a lower-mass star; and, for several species, the production by spallation processes in the winds of the forming Sun, for which see ~\citet{shu}. Uncertainties in the interpretations still remain, and
are due to the poorly known details of stellar nucleosynthesis and
of magnetic winds in star-forming regions.

The isochrons constructed thanks to short-lived radio-nuclei thus became a precious tool for estimating Myr-scale age differences among samples that
are, on average, 4.5 Gyr old. This offers a clock, with an accuracy that might be as high as 0.1 Myr, for dating the fast events that drove planetary formation \citep{mcd}.
For a more detailed account of the methods discussed in this section and of the science built in the last 50 years with short-lived nuclei I refer the reader to dedicated reviews -- see e.g. \citet{ap99,bus99,kratz,mcd,g+05,was06,wd07}. At the moment of this report, perhaps the most complete and recent update can be found in \citet{hus09}.

\section{Expected Conditions in the ESS}
\label{sec:6.3}
The presence, in the ESS, of short-lived radioactive nuclei, some of which exclusively produced
in stars, calls for a detailed study of the mutual relations between the forming sun and the
sources of galactic nucleosynthesis. In order to do this one needs to know the general scheme
of low-mass star formation, then inquiring whether some special condition was necessary, for
the sun to carry live nuclei of short mean-life. The conditions prevailing in star-forming
regions have been ascertained in the last 30 years, mainly thanks to the advances in the
techniques for infrared-, millimeter- and radio-astronomy.
Recent accounts can be found in \citet{lala} and in \citet{zuso}. Some 70-90\% of stars are born in clusters
or multiple systems, while the remaining part undergoes more isolated processes of slow accretion.
In the first phases not only gravity, but also galactic magnetic fields, velocity fields of the interstellar medium (ISM), and possibly the extra-pressure exerted by supernovae and other triggering phenomena can affect the proto-stellar clouds \citep{bos}.
In most cases, and especially when stars are born through the formation of isolated cloud cores or
globules, the initial phases are characterized by a slow accretion of cool materials thanks to
plasma processes, namely the diffusion of positive and negative particles at virtually the same rate,
due to interactions via the electric
fields. This mechanism, called ``ambipolar diffusion'' allows weakly ionized or neutral gas to separate from the
galactic plasma and to accumulate until the minimum mass necessary for gravitational contraction (the so-called
Jeans' mass) is reached. Then the collapse starts from inside out and a protostar and a disk are formed: see e.g. \citet{sh87a,sh87b} and references therein.

\subsection{Processes in Star-Forming Clouds}
\label{sec:6.3.1}
The various phases that lead a forming star to reach its final structure and burn hydrogen on the Main Sequence correspond to different classes of objects seen in molecular clouds at long wavelengths. Starting from a proto-stellar cloud (often indicated as being of class -I), one has first the formation of a condensed core with a surrounding envelope (class 0). This core then evolves, with the growth of the central condensation, through classes I, II and III, characterized by a varying spectral energy distribution at mid-infrared wavelengths, informative of the amount of cold circumstellar dust. Class I objects have large mid-infrared excesses and are optically invisible, Class II sources have infrared fluxes decreasing for increasing wavelength, as the percentage of dust is reduced to some $\sim$ 0.01 M$_{\odot}$. In the total emission, a flattened disk becomes gradually more important than the outside fading envelope and the central star becomes visible in the form of a variable object, of the T Tauri group (a star approaching the Main Sequence). Class III sources, then, have essentially no remaining mid-infrared excess from the original envelope, being``naked'' T Tauri stars, with disks whose masses decrease in time, although they are partially preserved up to the Main Sequence phase. The whole duration of the above processes is highly uncertain theoretically and also strongly dependent on the total mass. For solar-like stars the whole pre-Main Sequence evolution may last for about 20 Myr, half of which spent in the form of a pre-collapse cloud core. Then the formation of the central protostar is fast (virtually in free-fall conditions), and the T Tauri star may need another 10 Myr to approach the main Sequence and start the fusion of hydrogen into helium.

The above mentioned time scales suggest that the proto-sun must have been isolated from galactic nucleosynthesis processes for a time of one to several $\times 10^7$ yr.
Such a long quiescent time would imply large dilution
factors and long free-decay periods for any radioactive nucleus created in a galactic evolution scenario.
Roughly speaking, one might expect that nuclei with meanlives longer than about 10 Myr ($^{244}$Pu, $^{129}$I,
$^{182}$Hf etc.) might have survived the pre-stellar phases. On the contrary, other short-lived nuclei, with mean-lives up to a few Myr, would require a ``local'' production. With this term one may mean either that some nearby stellar source contaminated the ESS during its initial evolutionary stages, or that radioactive nuclei were produced
in the solar nebula itself. Very short-lived nuclei  will have to be attributed to either of these local processes; an exception to this rule is $^{53}$Mn. In fact, despite  its rather short lifetime ($\tau = 5.3$Myr), its production in supernovae is so huge that the requirement, for it, of a late addition has been questioned \citep{was06}.

Concerning the local production of radioactive nuclei, the stellar origin was early proposed by \citet{ct77}, who suggested that a nearby supernova might have been responsible both for introducing unstable nuclei in the ESS and for triggering the same contraction of the sun. An alternative to this assumed stellar event would require that, in the ESS, a combination of magnetized winds and particle bombardments might have induced spallation processes on early solids, breaking heavy nuclei into lighter unstable isotopes.

\begin{figure}[ht!!]
\begin{center}\includegraphics[width=0.8\columnwidth,clip]{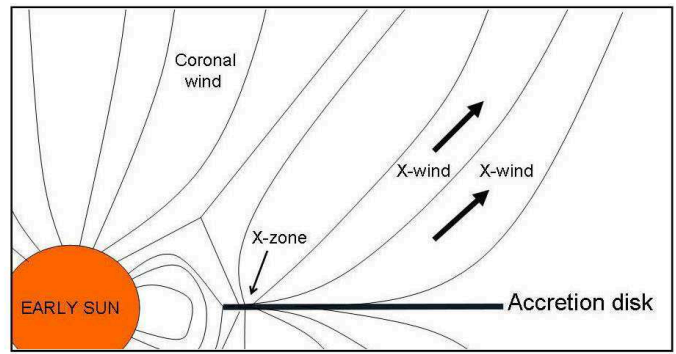}
\caption{A schematic view of the X-wind model}\label{fig:four}
\end{center}
\end{figure}

This last idea descends from the observation of strong bipolar emissions from star forming clouds, showing that matter is lost along the rotational axis through magnetic winds \citep{sh87a,sh87b,bos}. The physical origins of these winds have been identified in three main phenomena, namely: i) a coronal wind, originating directly from the star; ii) a disk wind, starting from the surface of the accretion disk over a wide range of distances from the central star (from less than one to more than a hundred astronomical units), which is probably the main source of mass loss in T Tauri variables \citep{hk96}; and iii) a wind driven by the interaction of the stellar corona with the inner edge of the accretion disk, launched at a distance of a few stellar radii from the center. This last is the so-called "X wind" \citep{shu1}, illustrated in Figure \ref{fig:four}, and is the type of magnetic interaction that might be important also for the explanation of extinct radioactivities in the solar nebula. In such a case, its role would be that of lifting early solids from the symmetry plane of the disk, exposing them to high temperatures and to fluxes of energetic particles from the sun. The irradiated dust would then fall back on the disk itself at large distances from the internal regions where it had been originally created, preserving the record of new radioactive species produced by solar spallation \citep{shu1,bos}. Confirmations of these lifting and transporting processes recently came from the first results of the STARDUST mission \citep{br06}.

\subsection{General Constraints for a Stellar Encounter}
\label{sec:6.3.2}
Since the original proposal by \citet{ct77}, the idea of a close stellar encounter for explaining the presence in the ESS of radioactive nuclei with short mean-life
has evolved into different branches, and both massive and low/intermediate-mass stars have been considered as possible sources \citep{pn97,was98,bus99,mc,was06,gou06,tak08,hus09}.

Various objections to the trigger model have however been advanced,
on the basis of the growing knowledge of star formation \citep{evan}. In the violent scenarios of cluster formation, events of
triggering are known \citep{koth,ps00,zin02}, induced either by SN ejecta, or by massive
star winds from OB associations. In such an environment, the required distance
of a SN  explosion from the solar nebula  can be estimated from the dilution
needed to obtain the ESS abundances of short-lived nuclei, starting with
their yields in SN ejecta. Useful distances are from a fraction of a pc to
a few pc, depending on the SN model; see also \citet{camal}. However, both the models for
the evolution SN remnants and the observational evidence indicate
that star formation triggering occurs when the shock fronts have slowed down to a few
km/sec or less \citep{pz99}. Such slow motions may require that the expanding SN shells have traveled much larger distances, of the order of 50pc: see e.g. \citet{se87}, chapter 6. This would imply a dilution in excess by a factor of about 10 as compared to the requirements by \citet{camal}. A SN trigger like the one originally suggested \citep{ct77}, if it ever occurred, must therefore be a very rare or unique event.

Recent studies have considered in some detail the probability that the Sun be formed in
a cluster massive enough to contain from one to several Supernovae, and that the timely explosion of these supernovae be at the origin of ESS radioactivities \citep{al01,hd05,wg07,sb07}. In such a case the contaminated nebula may have already evolved forming a protostellar disc \citep{al01,ou05,ou07}. A recent analysis in the above framework is  due to \citet{gm10}. For the contamination of a contracting nebula having already developed the protostar/disc structure these authors estimated the useful distance of the SN source to be very small (corresponding to less than 1 Myr
in time). It was also argued that massive-star evolution timescales are too long for
them to explode as supernovae with the due timeliness to contaminate forming stars in a
same cluster. As a consequence, in these works the probability of contamination from a
supernova belonging to the same cluster (the so-called Orion Nebula Cluster-like scenario) turns out to be very small (of the order of $10^{-3}$). Similarly low probabilities were
found by the above authors in the case of a more distant supernova, for a cloud still in
its initial evolutionary stages, those of a molecular cloud core, a situation that would mimic the original idea by \citet{ct77}.
If instead the Sun condensed alone, then a nearby supernova is even less likely then otherwise estimated.
One has further to remember that  all studies involving massive star evolution suffer for the large uncertainties involved in
crucial cross sections, especially the one for the $^{12}$C($\alpha,\gamma$)$^{16}$O reaction. A recent careful analysis
of SN yields for $^{26}$Al and $^{60}$Fe by \citet{tur09} concludes with warnings about the still non-quantitative nature of such
predictions.

For some radioactive nuclei an alternative was suggested by \citet{was94,was06} and by \citet{bus99,bus03}. This idea foresees the close-by passage of an AGB star in its final evolutionary stages, producing $^{26}$Al, $^{41}$Ca, $^{60}$Fe, $^{107}$Pd and $s$-process radioactivities like $^{135}$Cs and $^{205}$Pb. For updates on the production of some radioactive nuclei in AGB stars of intermediate mass see \citet{tr09}. However, also the close encounter with an AGB star is hard to motivate
on statistical grounds. In this case we have not to face the risk of cloud disruption (as for the fast supernova winds), but we require the simultaneous occurrence of two rare
phenomena (an isolated, low-mass, cloud-core collapse and a planetary nebula
ejection by an old star passing there just by chance). Critical remarks on this
idea, pointing out that such an event must be extremely rare, were advanced by
\citet{km94}. One has however to warn that most these remarks have been recently considered as being no longer valid \citep{tr09}. This is due to recent upgrades in our knowledge of the Initial Mass Function and also to the very poor statistics available at the time of the analysis by \citet{km94}.

 Anyhow, in order to overcome the difficulties related to the low probability of suitable encounters,
and of supernovae in particular, very recently the hypothesis was advanced that the presence
of $^{26}$Al in the oldest Solar System solids and its uniform distribution could be due
to the inheritance of $^{26}$Al itself from the parent giant molecular cloud, where
$^{26}$Al would have been contributed by very massive stars evolving to the Wolf-Rayet stage
\citep{gaidos2}. Although this idea is interesting and has now good reasons for being
further analyzed by other authors, in its first formulation it again leads to the conclusion
that the presence of $^{26}$Al is not a common event, and that the probability of this occurrence
is rather low. Moreover, ad hoc independent hypotheses had to be advanced to avoid overproduction
(by a factor of 10) of $^{60}$Fe from the supernovae expected in the cloud. The {\it combined} probability
of these special occurrences for $^{60}$Fe and of the production of $^{26}$Al by WR stars would certainly be extremely small. I guess that, needing two series of different (rare) events, the
joint explanation of both $^{26}$Al and $^{60}$Fe in this scenario would turn out to be even less likely
than for a single encounter (either a Supernova or an AGB star).

The emerging situation is very embarrassing. A few radioactive isotopes present alive in the ESS seem
to require necessarily a stellar production very close in time and space to the Sun's formation.
However, realistic (although highly uncertain) models of how
such a production might have occurred, always lead to unlikely conditions. It is not comfortable, for science, to call for special or unique conditions, occurred only for our sun. Somehow, this seems to shed doubts on the accepted principle, according to which our position in the Universe has nothing special. However, a more satisfactory view of the injection of radioactive nuclei into the ESS is not available at the present moment.

\section{The Galactic Inheritance}
\label{sec:6.4}
As mentioned, for radioactive nuclei with sufficiently long meanlife ($\tau \ge 10$ Myr)
the ESS abundance was probably inherited from the equilibrium established in the ISM by galactic evolution. This topic is dealt with extensively in Chapter 7; see also Chapter 2 for the effects on the ESS. For illustration purposes, we shall limit the discussion to the simplest possible scheme, representing the galaxy as a closed box with instantaneous recycling, evolving for a time interval $T$. For this case the equilibrium ratio $N_P/N_R$ of a radioactive nucleus $P$ with respect to a stable reference isotope $R$ produced in the same process was derived by \citet{sch}:

$$
N_P(T)/N_R(T) \simeq P_P \cdot p(T)\tau_{\rm
P}/P_R <p>T. \eqno (3)
$$

Here $P_R <p>$ is the stellar production rate of the stable reference
nucleus, expressed as the product of an invariant stellar
production factor $P_R$ and of the average over $T$ of a
variable scaling factor p(t). $P_P$ is the stellar
production factor for the parent isotope and $p(T)$ is the
scaling factor for it at time $T$. If the ESS is separated from
the ISM for a time interval $\Delta_1$ before forming the
first solid condensates, then the abundance ratio in the oldest meteoritic
material will be decreased with respect to the ISM equilibrium by
exp($-\tau_P/\Delta_1$). This old material is usually identified
with CAIs. Any younger solid body, formed at a time $t = \Delta_2$ after
the CAI's condensation, will have an abundance ratio further reduced
by exp($-\tau_P/\Delta_2$). In this framework, \citet{was96}
demonstrated that $^{244}$Pu (an actinide nucleus produced by the $r$-process and with a
relatively long mean life) and several other nuclei have ESS abundances compatible with
uniform production (hereafter UP) in the Galaxy over about 10$^{10}$ yr. On this subject, see also \citet{cam93,camal,mc}.

We already stated that the solar nebula must have been isolated from the last events of galactic nucleosynthesis for a time  $\Delta_1 > 10^7$ yr (see section \ref{sec:6.3.1}),
i.e. the time required by a cloud core to evolve. For reproducing the ESS amounts of  $^{182}$Hf this delay was found by ~\citet{was96} to be adequate. That scheme, however, would overestimate the concentration of $^{129}$I (and also of $^{107}$Pd) by a large amount, so that the above authors assumed that these isotopes were synthesized by a different supernova type, presenting an $r$-process distribution substantially different from the typical producers of heavy $r$-nuclei and whose last explosion would have occurred long before the one accounting for these last.

More specifically, assuming $r$-process production factors that were standard at that
moment, the above authors confirmed that reproducing $^{129}$I would require
a delay of about 70 Myr. This fact stimulated a series of works, both theoretical and
observational, on the ensuing multi-modal nature of the $r$-process. On these issues the constraints on the $r$-process coming from ESS radioactivites found complementary information from early stars in the galaxy, where the distribution of nuclei beyond Fe is best studied. Both phenomenological models \citep{qw07,qw08} and full $r$-process calculations in the so-called High Entropy Wind (HEW) scenario \citep{far06,far08,kra07} contributed to a lively discussion. At the moment of writing this report it appears that
the original proposition of (only) two distinct $r$-process components, one producing
$^{129}$I and the other producing actinides and $^{182}$Hf \citep{was96,qw} should be evolved into a more sophisticated picture of fast neutron captures, coming from a superposition of many distributions, all possibly produced in HEW environments, but each characterized by its own entropy range. Then the formation of the solar system would retain the signature of a special blend of distributions, the one characterizing the last $r$-process-producing episodes. In this view ESS radioactivities would in fact suggest a multi-modal $r$-process, as noted by \citet{was96}; but this would no longer lead to the possibility of inferring universal rules from the particular sample of supernova products experienced by the early Sun; see also, on this, \citet{kra08}.

The complexity of the situation is now confirmed by recent evidence on the $^{247}$Cm abundance in early solids. Until a few years ago, only a broad upper bound existed for this nucleus: $(^{247}$Cm/$^{235}$U)$_{ESS} << 2 \times 10^{-3}$, as established by \citet{cw81a,cw81b}. Subsequently, \citet{stir06} reported a lower bound in this ratio, close to 10$^{-4}$. Finally, unambiguous evidence of live $^{247}$Cm in ESS, at an abundance ratio $(^{247}$Cm/$^{235}$U)$_{ESS}$ = 1$-$2.4$\cdot10^{-4}$ has been found
by \citet{bren2}. This datum has two main consequences. First of all, as already commented upon, it shows that the ancient inventory of $^{235}$U might have been modified by actinide decays and should be corrected accordingly to avoid errors in the U-Pb datation system. On the other hand, the low $^{247}$Cm abundance inferred confirms that fast neutron-capture processes must be more complicated than assumed in the simple UP models, adopting ``standard'' $r$-process production factors. If such models are calibrated to reproduce the ESS concentration of $^{182}$Hf, then not only the predicted abundance of  $^{129}$I, but also that of $^{247}$Cm (and, to a lesser extent, of $^{244}$Pu) turn out to be definitely higher than observed. These facts would then require a significant time interval ($>$100 Myr) between the termination of actinide nucleosynthesis and the formation of the solar nebula; but this kind of nucleosynthesis would not produce $^{182}$Hf.

In the quoted works by \citet{was96,was06} it was suggested that, fixing the isolation time of the solar nebula to 70 Myr, so that the abundance of $^{129}$I predicted by UP models would fit the observed ESS concentration, then the inherited quantity of  $^{107}$Pd would turn out to be insufficient to explain its measured meteoritic abundance. As $^{107}$Pd can be produced also by the $s$-process in AGB stars, it was then natural to conclude that a last contamination phenomenon had occurred, induced by the close-by passage of an AGB star in its final stages of evolution, producing Pd but also $^{26}$Al, $^{41}$Ca, $^{60}$Fe, and possibly $^{205}$Pb.

The above approach was recently reconsidered by \citet{gm10}, who
noticed that, reducing the isolation time to 43 Myr, the discrepancies of the predictions for iodine and palladium abundances with respect to the measured requirements would be reduced to a factor of 3 (iodine being overabundant, palladium deficient, so that the discrepancy {\it between} the two is actually a factor of 10). According to those authors discrepancies at such a level might still be acceptable, in view of the large uncertainties affecting both models and observations, so that both $^{129}$I and $^{107}$Pd might be considered as products of the same continuous nucleosynthesis phenomenon in the Galaxy. One can however remember the old joke on the interpretation of experimental results, saying that: "agreement within a factor of ten" actually means "disagreement". The recent discoveries on the complexity of the $r$-process and the measurement of the $^{247}$Cm abundance confirm that this interpretation is not correct.

As mentioned, the above difficulties might be due to an insufficient picture of the $r$-process. Recent research in this field by \citet{far08} and \citet{kra08} suggests that a real, fast process of neutron captures is only responsible for nuclei above $A \ge$ 120, so that attributing $^{107}$Pd to the same process that produced $^{129}$I is probably wrong from a basic nucleosynthesis point of view. Below $A = $ 120 a blend of different nucleosynthesis mechanisms should be active, possibly induced by fast charged-particle processes, not by a "weak" $r$-process, as previously assumed \citep{far08}. Should these emerging conclusions be confirmed, both the schemes by \citet{was06} and by \citet{gm10}
would require revisions. Unfortunately, no general prescription
accounting completely for solar abundances in the mass range $A = 100 - 120$ is still
available. Contributions from a single neutron-burst, like the one ensuing from the passage of a SN shock through
the He-rich layers of the stellar mantle \citep{ma06}, are also possible, with yields for some heavy radioactive
nuclei hardly distinguishable from those of the r-process \citep{hus09}.
In such a condition, a late production of $^{107}$Pd can neither be securely
affirmed, nor securely denied.

\section{Local Production of Radioactive Nuclei}
\label{sec: 6.5}
We now consider those nuclei that cannot be ascribed to normal galactic evolution, as their very short meanlife would let them decay completely in the time delay required by a cloud core to evolve. For some of them (e.g. $^{60}$Fe) spallation processes cannot be at play, so referring to a nearby star seems unavoidable, despite the already mentioned low probability of such an event.

In the case of a close encounter, the relations that short-lived nuclei should obey at t=$\Delta_1$, were discussed by Wasserburg et al. (1994, 1998):
$$
\alpha_{R,S}^{\Delta_1} = ({N_R \over N_S})_{\Delta_1} \simeq
q_R(w)\cdot {N_R(w)\over N_S(w)}f_0 \cdot
e^{-{\Delta_1\over\tau_R}} \eqno (4)
$$
where $\alpha_{R,S}^{\Delta_1}$ is the abundance ratio
(radioactive nucleus to stable reference) established in the
solar nebula at t=$\Delta_1$, $q_S(w)$ is the production factor
of the stable nucleus $S$ in the stellar wind, $N_R(w)/N_S(w)$ is
the abundance ratio (radioactive to stable) in the wind, $f_0$ is
the dilution factor at t=0 and $\tau_R$ is the mean life of the
nucleus R. Abundances in PDs (t = $\Delta_2$)
can then be found by considering an exponential decay for
the nuclei of interest.
According to recent results from absolute dating of ESS samples \citep{zg02,am05} the values of $\Delta_2$ should not exceed few Myr.

\subsection{Short-Lived Nuclei: A Nearby Supernova Origin?}
\label{sec:6.5.1}

More than a decade ago  \citet{was98}, from an inspection of SN yields by \citet{ww95}, indicated
a fundamental constraint that must be satisfied by a SN explosion (at a time t=$\Delta_1$
before the formation of CAIs), providing short-lived nuclei in the ESS.
This constraint consists in ascertaining that the admixture of SN ejecta
with the material of the forming solar nebula be compatible with the
general knowledge of isotopic anomalies on stable elements in meteorites.
The main doubt that was raised in that paper is that a close SN would introduce
several anomalies over the distribution of stable isotopes and mainly on
typical SN products like oxygen and $\alpha$-rich nuclei, at levels that
should be within the present experimental possibilities. These problems
were underlined and repeated in great detail by \citet{was06}, considering
also models by \citet{rau02} and by \citet{lc03}. A remarkable anomaly, known
since several years, concerns $^{16}$O \citep{clay73,clay04}. However, it is not associated to
shifts in other oxygen isotopes \citep{th83,mau99} and it is now attributed to chemical
processes which do not involve any nucleosynthesis input. In this context, \citet{y+09} recently noted that the $^{18}$O/$^{17}$O ratio is anomalous in the
solar system as compared to the ISM. This anomaly was interpreted as due to the
addition of SN-polluted material to  the solar nebula, from various recent SNe in a sequential
picture of star formation. I notice however that this solar anomaly with respect to the present
ISM, if confirmed, can certainly be a signature of nucleosynthesis, but not necessarily of a
solar contamination by a SN. Addition of material richer in $^{17}$O than in $^{18}$O in the recent
evolution of the Galaxy, from low mass stars experiencing extra-mixing \citep{nol,pal09},
would induce the same difference, which should in this case be interpreted
as an ongoing production of $^{17}$O (and destruction of $^{18}$O) in the 4.56 Gyr after
the Sun's formation. This would underline the relevance of low mass star nucleosynthesis
in changing the chemical composition of objects younger than the Sun. This is a characteristic of recent galactic evolution that is now known from other contexts
\citep{dor09}. Very recently, conclusions similar to those discussed here (although not involving
extramixing) were presented by \citet{gaidos1}, who also pointed out how this scenario could explain
the gradual decrease in the $^{12}$C/$^{13}$C ratio of the Galaxy.

The SN models have been enormously improved since the time of the analysis by \citet{was98},
but the mentioned problem of their introduction of unwanted anomalies still remains.
Any SN explosion model that has to account for short-lived nuclei in the ESS should come
with a demonstration that it also does not imprint in the solar nebula signatures that are
incompatible with the known measurements. In view of the uncertainties discussed by \citet{tur09}
this seems hardly possible today.

With this caution in mind, we must recall that a wealth of new results have been
recently presented, both in modeling massive star nucleosynthesis/explosion
and in analyzing the possible pollution of the solar nebula from a supernova
\citep{cl04,mey05,lc06,nom06,hus07,wh07,tak08,kur,hus09}.
The above researches and the general requirements coming from ESS abundances
(e.g. avoiding the overproduction of $^{53}$Mn and of $^{60}$Fe as compared to $^{26}$Al,
guaranteeing enough production for very short-lived nuclei like $^{41}$Ca, etc)
focused the attention on SN models foreseeing the fallback of a consistent
amount of matter and including internal mixing among the previously layered composition
\citep{jog,tak08}. In particular, \citet{tak08} show that non-modified SN models by virtually all authors would overproduce $^{53}$Mn by large amounts, and most models would also
overproduce $^{60}$Fe. On the contrary, a "faint supernova", including
fallback and mixing, has much better possibilities to reproduce the observed distribution
of short-lived radioactivities. The mass cut below which material has to fall
back and the level of mixing among the stellar layers must then be varied as
free parameters to fit the measurements in the ESS. We recall that the first idea of
a modified supernova model with a high mass cut (to avoid $^{53}$Mn
overproduction) was suggested by the Clemson group \citep{mc}.  The models by
\citet{tak08} show now an impressively good accord between the ejecta of
a fall-back SN around 30 \Msol and the ESS record of $^{26}$Al, $^{41}$Ca, $^{53}$Mn and $^{60}$Fe.
A slight deficit in $^{53}$Mn from this model is not a real problem, in view of the abundant
production of this isotope in the continuous galactic nucleosynthesis processes.

Comparisons between the recent works by \citet{lc06} and by \citet{wh07} also clarify that accounting for the galactic inventory of $^{26}$Al~ and
$^{60}$Fe (for any reasonable value of their abundance ratio in the range indicated by
measurements)  is possible, provided the proper mass loss law is chosen. In this
respect the stronger mass loss rates by \citet{lan89} seem to be preferable to the more
recent choice by \citet{nl00}. The reproduction of the observed $^{26}$Al/$^{60}$Fe~ in
the galaxy is in any case an important point and if a massive star could also
explain the ESS radioactivities without introducing anomalies incompatible with
known meteoritic abundances, this would certainly offer a convincing global picture,
coming from a unique astrophysical site.

Summing up, at the present moment the scenario for a late supernova production of short-lived
nuclei is still the subject of an intense debate, in which arguments both in favor and against
this possibility are discussed. Among the reasons that make SNe appealing candidates one
can remember (at least) the following points:  i) SNe are in general the site where most
nucleosynthesis processes occur.
ii) A close-by SN might explain the abundances of ESS radioactivities in the same
framework in which the equilibrium abundances of radioactive nuclei in the Galaxy
are explained. iii) Despite the difficulties we mentioned, star formation in the vicinity of
SNe and possibly triggered by them is known to occur in various galactic environments.
iv) Faint supernovae, with internal mixing and limited or no contribution to iron
do exist, so that the speculative part of the models required by ESS radioactivities is
at least plausible.

However, serious concerns arise by considering other relevant issues, such as:
i) The fact that the probability of a close-by SN encounter has been recently shown to be extremely
small \citep{gm10}, possibly smaller than for an AGB star \citep{km94,tr09}.
ii) The need for ad-hoc choices of parameters, like e.g. the mixing extension and the mass cut,
in any SN model suitable to explain the ESS radioactivities and the fact that results strongly depend
on uncertain choices for cross sections \citep{tur09}. iii) The lack of a convincing answer
to the risk of introducing unwanted anomalies on stable isotopes. As an example, the best model by \citet{tak08},
from a 30 $M_{\odot}$~star, ejects almost 7 $M_{\odot}$~of processed material, where at least C, O, Ne, and Mg
should be highly enhanced (maybe by factors close to 100, as other SN nucleosynthesis calculations suggest).
At the adopted dilution factor of 4$\times$10$^{-4}$ this should induce anomalies at percent level on major
elements, which are not observed. This inconsistency should be disproved before any model involving a SN can be considered as a real, quantitative possibility.

\subsection{Contributions from a Nearby AGB Star?}
\label{sec:6.5.2}
As already mentioned, the scenario of a close-by contamination by an AGB star
has been extensively explored in the past \citep{was94,was95b,was06}. A recent detailed analysis of the contributions from
an intermediate mass star experiencing H burning at the base of the envelope (Hot Bottom Burning, or HBB) was presented by \citet{tr09}; these authors also demonstrated that previous criticisms the the AGB scenario might be inconsistent, or at least should be looked at with a lot of caution.

The most appealing property of the above AGB models is that they don't need
to fix free parameters that are not otherwise constrained. As an example,
neutron capture nucleosynthesis at the relatively low efficiency
required to explain the ESS radioactivities does not need neutrons from
the \ct~ neutron source, which still lacks a first-principle modeling. Hence
the only process still needing parameters not provided by the main stellar
evolution history is the efficiency of extra-mixing processes above the H-burning shell, which are necessary to produce $^{26}$Al~ if the mass of the star is too small to host HBB (i.e. lower than 4 M$_{\odot}$). The existence of thermally-pulsing AGB stars with minimum
$s$-process production, so that a \ct-pocket becomes unnecessary, was known already
for low mass stars (the Mira protoptype, $o ~Cet$ being an example) and has been recently
demonstrated also for intermediate masses \citep{gh07}.

The effectiveness of extramixing can be independently established, for low mass stars, from consideration of abundances in presolar grains of AGB origin, whose reproduction
offers a way to constrain in a rather clear way the occurrence of any non-convective
circulation of matter.  One has also to notice that, although a unique paradigm for
extramixing is still lacking, recent work may have provided new tools to solve this
old problem. \citet{egg1} and \citet{egg2} have underlined the role of instabilities (thermohaline mixing) induced by the molecular weight decrease associated to the burning of two $^{3}$He nuclei, producing $^{4}$He and two protons. Alternatively, a similar action can be provided by magnetic fields; in their flux tubes the gas density is smaller than in the outside environment, so that the buoyancy of magnetized blobs, processed by proton captures, can occur \citep{bus07,nor,den}. In any case, as shown by \citet{nol} and by \citet{pal09}, the actual extra-mixing products do not in general depend on the specific physical model accounting for it (exception however exist, the most remarkable one being offered by lithium).

The mentioned minimum use of parameters might be even more true for an intermediate mass star, where H-burning at the base of the convective envelope provides the high-temperature environment where to produce $^{26}$Al~ and non-convective mixing mechanisms are not required (and actually not expected to occur).  In general, both low and intermediate mass stars will need a free choice for mass loss rates, which only now start to be quantitatively estimated, especially thanks to extensive sky surveys in the infrared by space-borne instruments \citep{gua06,bus07a,gub08}.

A crucial point for AGB models is the ESS abundance of $^{60}$Fe. A low mass star ($M = 1.5 - 2$ \Msol) would be incapable of producing enough $^{60}$Fe~ to account for the initial ESS $^{60}$Fe/$^{56}$Fe ratio (at least a few$\times$10$^{-7}$). Hence the calculations by both \cite{was06} and \citet{tr09} pointed toward stellar masses of 3-7 M$_{\odot}$, higher than in the original suggestions by \cite{was94}. These recent works show that both a 3 \Msol~ and a 6.5 \Msol~ star can reproduce well the ESS measurements at least for $^{26}$Al, $^{41}$Ca, $^{60}$Fe, and $^{107}$Pd, for dilution factors of a few $\times$10$^{-3}$. This is shown in Table 3, where the best cases from  \cite{was06} and \citet{tr09} are summarized; note that the estimates for $^{135}$Cs and $^{205}$Pb are available only for the results of \cite{was06}. In the 3 M$_{\odot}$ model $^{26}$Al  would come from extra-mixing, in the 6.5 M$_{\odot}$ star from HBB. We must also underline that AGB stars of intermediate mass would imprint in the solar nebula only marginal anomalies in stable nuclei. The most critical effect is probably a 1\% shift in $^{17}$O for a star of 6.5M$_{\odot}$, see \cite{tr09}. A lower-mass star would have essentially no effect on stable isotopes, except carbon. This is a special merit of AGB stars. If one also considers that heavy $s$-process nuclei like $^{205}$Pb are typical products of AGB nucleosynthesis, some cautions should be used before excluding an intermediate-mass star as a polluting source, at least until massive stars cannot be shown to provide the same consistency. Very recently a remarkable nuclear physics result has added new arguments in this direction. The half-life of $^{60}$Fe was remeasured by \cite{rugel} and the new value is about a factor of two larger than
the one previously adopted. If this will be confirmed, the main challenge for AGB models (that of producing enough $^{60}$Fe) would be considerably alleviated
and AGB nucleosynthesis would offer an even more promising scenario than so far suspected.  This scenario does not suffer for the same nuclear uncertainties
discussed for massive stars and, despite all contrary arguments, it remains the only really quantitative model on the market, thanks to its relatively simple physics.

One must however recall (as already mentioned in section \ref{sec:6.3.2}) that strong doubts on the possibility of an AGB pollution were
advanced in the past by \citet{km94}, estimating to a mere 1\% the probability of a close encounter. As mentioned, recently \citet{tr09} presented
several reasons in favor of a revision of this estimate. In any case, the recent discussion by \citet{gm10} seem to indicate that encounter probabilities for massive stars are actually at least as small as for AGBs. If a close encounter with a star producing radioactive nuclei ever occurred, this must have been necessarily an unusual event, maybe a unique occurrence. If this is the case, then the relative statistics is no longer an issue in judging these scenarios.

An important role in deciding between an AGB star and a massive star as a source for ESS pollution would be a better knowledge of the initial concentration of $^{205}$Pb. Among heavy, $n$-capture nuclei, $^{135}$Cs can be produced by both the $s$- and $r$-processes, while $^{205}$Pb is a purely $s$-process nucleus. At the low neutron exposure required, AGB stars do not produce it much more efficiently than a massive star (we are in a situation very different from the main component of the $s$-process, which is instead the exclusive realm of AGB stars). However, if $^{205}$Pb~ was really present in the ESS at the level suggested by the measurements of \citet{niel}, then a massive star would be essentially excluded for a question of dilution factors.
 For a dilution of some 10$^{-3}$, \cite{was06} obtained an ESS concentration of a few $\times$10$^{-4}$ for $^{205}$Pb, in the case of AGB production. This has to be interpreted as an upper limit, given the uncertainties affecting  the decay rate of $^{205}$Pb in the stellar interior. Looking at recent results from \cite{wh07}, $^{205}$Pb~ appears to be produced by a 25\Msol~ star more or less at the same efficiency as for an AGB star of intermediate mass, but the material would then need to be diluted 10 times more, to avoid overproduction of lighter species. This is still a very difficult matter to put in focus, because of the large uncertainties affecting both the measurements and the predictions. We underline this point here because it deserves further attention. It would in particular be very important if theoretical nuclear physics could firmly establish the complex behavior of the weak interactions regulating the pair $^{205}$Pb-$^{205}$Tl as a function of temperature, in the range between, say, 10$^6$ and 10$^{8}$ K  (pertinent to the stellar layers that any $^{205}$Pb~ produced by neutron captures must cross while being dredged-up to the surface of an AGB star, or ejected in a SN explosion).

\begin{table}[h!!]

\begin{center}

{\bf TABLE 3. \\
Recent Predictions for Short Lived Nuclei from an AGB Star.

\vspace{0.5cm}
\rm
\begin{tabular}{|c|c|c c|c c|c|}

\hline \hline
 & & & & & &\\
 & & 3.0 \Msol, $Z = Z_{\odot}/3$ & (W+06)  & 6.7 \Msol, $Z = Z_{\odot}$, & (TR+09) & Measured \\
 & & $f_0=4.0\times10^{-3}$& &$f_0=3.3\times10^{-3}$ &  & or extrapolated\\
 \hline
 & & & & & &\\
  Parent & Index &$(N_P/N_I)_{\Delta_1}$ & $(N_P/N_I)_{\Delta_2}$ &  $(N_P/N_I)_{\Delta_1}$ & $(N_P/N_I)_{\Delta_2}$ & at $t = {\Delta_i}$\\
 $P$ & $I$ & $\Delta_1$ = 0.53 Myr &$\Delta_2$ = 6.5 Myr &$\Delta_1$ = 0.53 Myr & (${\Delta_2}$) = 6.0 Myr &  \\
\hline
 & & & & & & \\
  $^{26}$Al & $^{27}$Al  & $5.0\cdot10^{-5}$ & $8.5\cdot$10$^{-8}$ & $3.2\cdot10^{-5}$&$9.8\cdot10^{-8}$ & $5.0\cdot10^{-5}$ (${\Delta_1}$)\\
 & & & & & &\\
  $^{41}$Ca & $^{40}$Ca    & $1.5\cdot10^{-8}$ & -- & $1.5\cdot10^{-8}$& -- & $\ge 1.5\cdot10^{-8}$ (${\Delta_1}$)\\
 & & & & & & \\
  $^{60}$Fe & $^{56}$Fe    & $2.1\cdot$10$^{-6}$ & $1.0\cdot10^{-7}$ &$2.6\cdot$10$^{-6}$ & $1.7\cdot10^{-7}$ & $0.5 - 1\cdot$10$^{-6}$ (${\Delta_1}$)\\
 & & & & & & \\
  $^{107}$Pd & $^{108}$Pd  & $3.8\cdot10^{-5}$ & $2.0\cdot$10$^{-5}$ & $3.8\cdot10^{-5}$ & $2.0\cdot$10$^{-5}$  & $2.0\cdot$10$^{-5}$ (${\Delta_2}$)\\
 & & & & & & \\
 \hline
 & & & & & & \\
   $^{93}$Zr & $^{92}$Zr  & $2.5\cdot10^{-4}$ & $1.2\cdot$10$^{-5}$ & $1.6\cdot10^{-4}$& $8.6\cdot10^{-6}$ & (?)\\
 & & & & & & \\
  $^{99}$Tc & $^{100}$Ru  & $1.9\cdot$10$^{-5}$ & -- & $1.4\cdot$10$^{-5}$& -- & (?) \\
 & & & & & & \\
  $^{135}$Cs & $^{133}$Cs & $3.6\cdot10^{-5}$ & $3.5\cdot$10$^{-6}$ & n.a. & n.a. & 1.6$\times$10$^{-4}$ (?)\\
 & & & & & & \\
  $^{205}$Pb & $^{204}$Pb & $\le 3.4\cdot10^{-4}$ & $\le 2.5\cdot$10$^{-4}$ & n.a. & n.a. & $1 - 2\times$10$^{-4}$ (?)\\
 & & & & & & \\
\hline\hline
\end{tabular}
}
\end{center}
\end{table}

Summing up, also the models of an AGB contamination contain merits and problems.
Among the merits one can recall that: i) The AGB phase is a experienced by all stars below about 7 $-$ 8 M$_{\odot}$, i.e. by the large majority of galactic objects evolving within a cosmological time scale. Possible candidates are abundant in any galactic epoch. ii) AGB models are simpler, involve a much smaller number of free parameters than for massive stars and their main features are not critically dependent on uncertain reaction rates. Most parameters can be fixed independently, from the abundant observations, and the predictions can therefore be rather quantitative. iii) A polluting AGB star of about 3 M$_{\odot}$(or even less, accepting the revision of the $^{60}$Fe lifetime by
\citet{rugel} would not introduce any unwanted anomaly on stable nuclei. iv) An AGB star would be able to produce also heavy nuclei like $^{135}$Cs and $^{205}$Pb, at concentrations close to those suggested by the present, very limited, data. Firm experimental constraints would however be needed.

Despite the above advantages, the AGB scenario still generates doubts and concerns, mainly due to the following problems: i) An AGB encounter is difficult to imagine for a cloud core; apart from the low probability derived by statistical calculations, a proximity of forming stars to AGBs or planetary nebulae is certainly not a common fact (if it can occur at all), while it is demonstrated for SNe. ii) A small dilution is required,
due to the small amount of mass ejected: this requires very tight constraints on the timing. iii) As nuclei deriving by the continuous galactic production are accounted for by SN nucleosynthesis, a late event of different nature appears rather ad-hoc.

\section{Short-lived Nuclei: In-situ Production}
\label{sec:6.6}
As already mentioned, for a few
shorter-lived species, especially $^{10}$Be, $^{26}$Al, $^{36}$Cl, $^{41}$Ca, $^{53}$Mn,
alternative models for the formation in spallation
reactions, from the bombardment of fast particles coming from the
magnetically-active early Sun were proposed; see e.g. \citet{shu1}; \citet{goun}.
The same magnetic phenomena occurring in most stars, at least during their Main Sequence
but probably also in their subsequent Red Giant phases \citep{andal}, provide a site where nucleosynthesis of radioactive isotopes occurs: this has been shown for very short-lived nuclei \citep{tati1} but might be true also for longer-lived species of interest for the ESS \citep{pb08}.

\subsection{Radioactivities from the Bombardment of Early Solids}
\label{sec:6.6.1}
This mechanism takes advantage of the already described X-wind scenario \citep{lee4}, lifting the recently formed CAIs from the disk plane and transporting them to large distances. In such models the hypothesis was also advanced that CAIs and chondrules
might be produced in those same winds. If this were the case, then the high abundance
measured in CAIs for $^{26}$Al~ ($^{26}$Al/$^{27}$Al = 5$\cdot10^{-5}$) would
not be indicative of a uniform situation, and many properties
attributed to $^{26}$Al~ would have to be reconsidered, especially its use as
a precise chronometer. A thorough discussion of the spallation mechanisms
in the ESS was presented by \citet{gou06}, on the basis of nuclear parameters for spallation processes that were standard at that epoch. On the basis of models for the structure of CAIs and for the flux of irradiating particles, the authors first determined the conditions for the production of $^7$Be, whose very short mean life makes it a discriminating nucleus. They suggested that, for a flux of particles from the sun of the order of $F \simeq 2\times10^{10}$ cm$^{-2}$ sec$^{-1}$, both $^7$Be and $^{10}$Be could be produced at the observed levels within the uncertainties. A byproduct of the adopted fluxes was also a noticeable contribution to $^{26}$Al, $^{36}$Cl, and $^{53}$Mn, whose abundances in the ESS were therefore suggested to come form proton bombardment, leaving space to a nearby supernova only to produce $^{60}$Fe. Further analysis lead \citet{g+09} to suggest that $^{60}$Fe itself
(at a level close to the lower limit of the uncertainty range for its abundance)
might derive not from an isolated, rare event, but rather from the superposition of various SN contributions, much like for longer-lived species.

On the above subject one can however notice that SN nucleosynthesis has a complex pattern: $^{60}$Fe cannot be produced alone. Other radioactive species would come out, and their predicted abundance in the ESS should be verified. The effects of several recent supernovae would also be detectable from many abundance anomalies in stable elements at levels within the reach of present measurements. This issue is the same previously raised by \citet{was98,was06} for one isolated SN.

After the work by \citet{gou06}, it was shown that the cross sections for some crucial spallation processes, in particular $^{24}$Mg($^{3}$He,p)$^{26}$Al, needed strong revisions \citep{fit}. This leads to the conclusion that $^{26}$Al cannot be produced at sufficient levels by spallation processes in the early solar nebula, so that its synthesis seems now to require unambiguously nucleosynthesis processes in stars. The clear decoupling between $^{26}$Al and $^{36}$Cl \citep{hsu} in early meteorites then suggests that $^{36}$Cl, instead, was formed in the the X-wind scenario, together with Be-isotopes. The already mentioned correlation of $^{41}$Ca with $^{26}$Al (see Figure \ref{fig:three} would instead suggest a stellar origin for $^{41}$Ca. Further limits to the level of production of radioactive nuclei from solid particle bombardment were put, after
the first results of the STARDUST mission \citep{br06}, by new models using improved cross sections for the reactions leading to $^7$Be, $^{10}$Be,
$^{26}$Al, $^{36}$Cl, $^{41}$Ca, and $^{53}$Mn \citep{dt08}. According to these new findings, the role of non-thermal nucleosynthesis in the ESS might actually be limited to the production of $^{10}$Be and $^{41}$Ca.

All these difficulties of the solar-wind models then call again for a last-minute stellar contamination, at least for $^{26}$Al, $^{41}$Ca, $^{60}$Fe, $^{135}$Cs, $^{205}$Pb. At this moment the need for an exceptional nucleosynthesis event of some kind, close in time and space to the forming sun, cannot be avoided. Although this event looks very unusual in the light of our understanding of star formation, we might be forced to accept that an unlikely event occurred at least once, if the alternative is no explanation at all.

\subsection{Solar Activity and the production of $^{7}$Be}
\label{sec:6.6.2}
Nuclear interactions occurring in solar explosions are revealed by their induced prompt
emission of $\gamma$-ray lines; these are due to the de-excitation of nuclei excited by reactions occurring between flare-accelerated particles and the solar atmospheric material. The first observations of these solar $\gamma$-ray lines were obtained by the experiment GRS (Gamma Ray Spectrometer),
on board of the OSO-7 space-borne observatory \citep{chupp}. Subsequently, the evidence for spallation-induced
reactions in flares grew thanks to the measurements of various instruments: the Solar Maximum Mission \citep{sm95},
the Compton Gamma-Ray Observatory \citep{sh97} and the Ramaty High Energy Solar Spectroscopic Imager (RHESSI), see e.g.
\citet{lin3}.

In recent years, it was suggested that also delayed X- and $\gamma$-ray emission
might occur from solar flares, thanks to the production of short-lived radioactive nuclei, whose subsequent decay
would be accompanied by emission lines \citep{rm00,koz02}. In particular, detailed predictions of line emissions
form such decays in solar active regions, after the occurrence of intense flares, were presented by \citet{tati1,tati2}.
These authors estimated the cross sections for the formation of several radio-nuclei from interactions of
protons and $^{3,4}$He particles with isotopes of elements up to Ni. Then, on the basis of a thick-target model,
they provided expected yields for 25 radio-nuclei with half-life in the range from 10 minutes ($^{13}$N) to
77 days ($^{56}$Co). Fluxes in $\gamma$-ray lines interesting for a possible detection from future experiments
were found for $^{52}$Mn (1434 keV), $^{60}$Cu (1332, 1792 keV), $^{34}$Cl (2127 keV), $^{24}$Na (1369, 2754 keV) and
$^{55}$Co (931.1 keV).

The above findings are important for explaining the $\gamma$-ray fluxes of solar activity phenomena and
for providing information on details of the solar plasma physics. On the contrary, spallation processes on solar
and stellar atmospheres are not expected to be relevant for galactic nucleosynthesis, as production of nuclei
by similar processes occurring in Galactic Cosmic Rays would dominate by several orders of magnitude
\citep{tati3}. Concerning the ESS radioactivities discussed in this Chapter, the works mentioned in this subsection
might be of interest for the early inventory of very short-lived nuclei, like $^7$Be. In fact, it has been found
that this isotope might have been present alive in CAIs \citep{chaus1,chaus2}. Its production must be
essentially contemporaneous to the same CAI formation, due to its very short half-life (53 days), and its (uncertain) ESS abundance was originally indicated as being roughly compatible with X-wind models \citep{gou03,chaus2}. Recent revisions suggest, instead, that its production in that environment should be largely insufficient \citep{dt08}. The formation of $^7$Be in solar flares might therefore be considered as a promising alternative possibility to the irradiation of solids. It has also been noticed
that production of $^7$Be (hence of its daughter $^7$Li) might explain the unexpected detection of Li in several M-type dwarfs, although it would probably be insufficient to account for the trends of the Li abundance in open clusters \citep{tati4}.

The above studies represent a remarkable bridge linking the models for the in-situ production of ESS radioactivities
to those for their stellar synthesis. Indeed, some recent models for the deep-mixing phenomena
occurring in evolved red giants and accounting, among other things, for the production of abundant $^{26}$Al and Li
\citep{pb08,pal09,gua09} consider stellar magnetic fields as the main engine for the transport of proton-capture
ashes to the stellar envelope, following ideas by \citet{bus07}. Those models account for the production and destruction of Li in red giants through
mechanisms of magnetic buoyancy occurring at rates respectively faster or slower than that for $^7$Be decay. Such models,
however, did not include, so far, the possible contributions from spallation processes in the transported material itself,
which is moving relativistically along toroidal flux tubes and their $\Omega$-shaped instabilities. A check of
the relevance of stellar spallation processes for explaining the formation of super-Li rich stars, and for
the evolution of the Li abundance in the Sun seems to be really necessary now.

\section{Conclusions}%
\label{sec:6.7}

The general scenarios explored so far in order to account for the presence of short-lived radioactive nuclei in the ESS is far from satisfactory and still quite confused.
We can summarize the indications emerged from the many efforts dedicated to this field in the past years in the following points:
\begin{itemize}
\item{Decay of radioactive isotopes of long lifetime (longer than 1 Gyr)
offers us a number of tools for estimating the global age of the
Earth and of the other solid bodies orbiting around our star. They tell us
that the solar system was formed when the Galaxy was already quite
mature, having spent 2/3 of its present age. The resulting age of the solar system
is close to 4.566 Gyr.}

\item{The ESS abundances of shorter-lived isotopes, with half lives from 10 to a few hundred Myr, can be used for reconstructing the history of
nucleosynthesis in the solar neighborhood, as their initial
concentration in the solar nebula is probably compatible with the
equilibrium abundances established by SN explosions and nuclear
decay in the local interstellar medium, during several cycles of
molecular cloud aggregation and destruction. From these nuclei it was inferred that the $r$-process has a multi-modal nature. This suggestion, which has produced considerable insight on both solar system formation and the abundance patterns observed in old stars, has stimulated further studies that are now producing a very complex scenario. First of all, it has been shown that some of the radioactive nuclei commonly ascribed to the $r$ process might be produced also in a single neutron burst in evolved massive stars. Secondly, the classical $r$-process itself has undergone important revisions,
and seems now to be limited to the production of very heavy nuclei (A $>$ 120).}

\item{The solar nebula must have undergone a process of autonomous evolution in which it was isolated from SN nucleosynthesis contaminations for a rather long time. One can reconstruct, from the initial abundance ratio of heavy n-rich species,  that the last episodes of SN nucleosynthesis contributing to these nuclei affected the solar material from 10$^7$ to 10$^8$ yr ago. More stringent constraints require a better knowledge of fast nucleosynthesis processes beyond Fe.}

\item{There are also signs that the forming Sun was affected by a last
episode of stellar nucleosynthesis (much closer in time than those mentioned above) producing some short lived radioactivities (like $^{26}$Al, $^{41}$Ca, $^{60}$Fe, and possibly $^{135}$Cs, $^{205}$Pb), although this event has a low probability of occurrence from a purely statistical point of view. The stellar contamination had to occur immediately
before the contraction of the solar parent cloud (with a time separation from it of $\Delta_1 \le$ 1$-$ 2
Myr). The importance of this last event was crucial for several reasons. In particular, decays from some radioactive nuclei generated in this event, like
$^{26}$Al~ and $^{60}$Fe, produced daughter isotopes in excited states. Their de-excitation is now considered as the main heating source for differentiating
the early bodies \citep{urey,sch}.}

\item{Attribution of this last event to a massive Supernova of some
peculiar type or to an AGB star is still a matter of lively debates, as both
scenarios face unsolved problems and have considerable drawbacks.}

\item{Subsequently, the forming Sun itself, in its fully convective, pre-Main
Sequence phase, must have added new nuclei (\be, $^{10}$Be, $^{36}$Cl) through spallation processed occurring either in coronal flares or in the interactions of the solar wind with early solids that were forming in the inner regions of an accretion disk.}
\end{itemize}

Despite the many uncertainties, the wealth of the short-lived or very-short-lived radioactivities discussed in this Chapter is now an invaluable source of information on the timing of the first events occurred in the solar nebula.

\bibliographystyle{spbasic}

\backmatter
\printindex

\end{document}